\documentclass[twocolumn,aps,amsmath,prd,amssymb,eqsecnum,nofootinbib]{revtex4-1}

\usepackage{graphicx}

\begin{document}

\title{Quantum field theory on the Bertotti-Robinson space-time}

\author{Adrian C. Ottewill}
\email{adrian.ottewill@ucd.ie}
\affiliation{School of Mathematical Sciences and Complex \& Adaptive Systems Laboratory, University College Dublin, Belfield, Dublin 4, Ireland}
\author{Peter Taylor}
\email{peter.taylor@maths.tcd.ie}
\affiliation{School of Mathematics, Trinity College, Dublin 2, Ireland}

\date{\today}
\begin{abstract}
We consider the problem of quantum field theory on the Bertotti-Robinson space-time, which arises naturally as the near horizon geometry of an extremal Reissner-Nordstrom black hole but can also arise in certain near-horizon limits of non-extremal Reissner Nordstrom space-time. The various vacuum states have been considered in the context of $AdS_{2}$ black holes \cite{StromingerAdSBH} where it was shown that the Poincare vacuum, the Global vacuum and the Hartle-Hawking vacuum are all equivalent, while the Boulware vacuum and the Schwarzschild vacuum are equivalent. We verify this by explicitly computing the Green's functions in closed form for a massless scalar field corresponding to each of these vacua. Obtaining a closed form for the Green's function corresponding to the Boulware vacuum is non-trivial, we present it here for the first time by deriving a new summation formula for associated Legendre functions that allows us to perform the mode-sum. Having obtained the propagator for the Boulware vacuum, which is a zero-temperature Green's function, we can then consider the case of a scalar field at an arbitrary temperature by an infinite image imaginary-time sum, which yields the Hartle-Hawking propagator upon setting the temperature to the Hawking temperature. Finally, we compute the renormalized stress-energy tensor for a massless scalar field in the various quantum vacua.
\end{abstract}
\maketitle

\section{Introduction}
The near-horizon limit of an extremal Reissner-Nordstrom black hole is described by the direct product spacetime $AdS_{2}\times\mathbb{S}^{2}$ \cite{Carter} known as the Bertotti-Robinson spacetime \cite{Robinson, Bertotti}. The $AdS_{2}\times\mathbb{S}^{2}$ geometry also arises in various other inequivalent near-horizon limits of the non-extremal Reissner-Nordstrom black hole \cite{MaldacenaStrominger}. In recent decades, interest in the Bertotti-Robinson space-time has gained impetus from a string theory perspective \cite{MaldacenaStrominger, StromingerAdSQG, StromingerAdSBH}, where anti-de Sitter black holes have played a crucial part in the AdS/CFT correspondence, particularly in the two-dimensional case \cite{CadoniMignemi1, CadoniMignemi2, CadoniMignemi3, CadoniMignemi4, CaldarelliVanzo, CatelaniVanzo, CadoniCavaglia, Navarro, CadoniMignemi5}. Of particular relevance to this paper is the work of Spradlin and Strominger \cite{StromingerAdSBH} who derive the two-dimensional propagators for various quantum states on the $AdS_{2}$ black hole space-time and use these results to compute the renormalized stress-energy tensor for massive and massless scalar fields. When discussing vacuum states on the Bertotti-Robinson space-time, one can ignore the angular degrees of freedom since the vacuum state is determined by requiring modes to be positive frequency with respect to a particular time coordinate. Hence, the quantum states will be equivalent to those on $AdS_{2}$. However, the physical quantities of interest on the Bertotti-Robinson space-time, such as the renormalized stress-energy tensor, involve the Green's function for the wave equation whose structure is very different in the two dimensional and four dimensional cases. We shall consider only the massless scalar field, the massive field propagator cannot be obtained in closed form but one can employ a large-mass approximation for the stress-energy tensor which does not depend on the global properties such as the quantum state. Such large-mass approximations have been adopted in the study of the quantum-corrected Bertotti-Robinson space-time (see, for example, \cite{MatyjasekTryniecki} and references therein).

The Green's function on the Bertotti-Robinson space-time was derived in closed form in Poincare coordinates by Kofman and Sahni \cite{KofmanSahni1}, which corresponds to the field in the Poincare vacuum defined by choosing modes to be positive frequency with respect to Poincare time. Spradlin and Strominger \cite{StromingerAdSBH} show that this vacuum is equivalent to the Global vacuum and the Hartle-Hawking vacuum and hence the Green's function for these states are equal. By definition, the Hartle-Hawking vacuum respects the isometries of the space-time and hence the corresponding propagator has a very simple structure, factorizing into a part that depends on the geodesic distance on $AdS_{2}$ and a part that depends on the geodesic distance on $\mathbb{S}^{2}$. The Green's function in the Boulware or Schwarzschild vacuum is more complicated than the Hartle-Hawking Green's function and to the best of our knowledge has not been obtained in closed form. We present a closed-form representation here, verifying our result by reproducing the known Hartle-Hawking Green's function by an infinite image imaginary-time sum.

The mode-sum representation of the Green's function in the Boulware vacuum involves a sum over associated Legendre functions of non-integer order. Performing this sum is crucial in arriving at the closed-form Green's function. The method employed is to associate the sum with a three-dimensional Green's function on a dimensionally reduced space-time which satisfies a Helmholtz equation whose zero-potential solution is known in closed-form. We then expand about the zero-potential solution which yields a closed form. This expansion is similar in spirit to Copson's method \cite{Copson} for obtaining the electrostatic potential in Schwarzschild space-time. The derivation of this result is lengthy so we defer it to the appendix rather than disrupt the continuity of the paper.

The layout of this paper is as follows: In Sec.~\ref{sec:rslimits}, we review the various limits of Reissner-Nordsrom space-time that result in the Bertotti-Robinson geometry. We also discuss the many coordinate systems that define the vacuum states. In Sec.~\ref{sec:propagators}, we derive the Feynman Green's function for several quantum states. There are essentially three distinct cases, the zero-temperature Boulware vacuum which we denote by $\big|B\big>$, the mixed state for a field at arbitrary temperature $T=1/\beta$ which we denote by $\big|\beta\big>$ and the mixed state for the field at the Hawking temperature, which is analogous to the Hartle-Hawking state in Schwarzschild space-time \cite{HartleHawking}, we denote it by $\big|H\big>$. Finally, in Sec.~\ref{sec:rset}, we obtain analytic expressions for the renormalized stress-energy tensor for a massless field in each of these states.

\section{The Bertotti-Robinson Space-time as a Limit of the Reissner-Nordstrom Solution}
\label{sec:rslimits}
In this section, we show how the Bertotti-Robinson geometry emerges as the near-horizon limit of the extreme Reissner-Nordstrom black hole and also in certain near-horizon limits of non-extremal Reissner-Nordstrom black holes. 

The Reissner-Nordstrom solution is the unique static, spherically symmetric solution of the Einstein-Maxwell equations. In units where the speed of light, Planck's constant, the Boltzmann constant and the Coulomb constant are set to unity, the metric is given by
\begin{equation}
ds^{2}=-\frac{\Delta}{r^{2}}dt_{s}^{2}+\frac{r^{2}}{\Delta}dr^{2}+r^{2}d\Omega_{2}^{2}
\end{equation}
where $\Delta=(r-r_{-})(r-r_{+})$ and $d\Omega_{2}^{2}$ is the metric on the two-sphere. We choose to denote the usual Schwarzschild time coordinate here by $t_{s}$. The Cauchy horizon $r_{-}$ and the black hole horizon $r_{+}$ are
\begin{equation}
r_{\pm}=L_{p}^{2} M\pm L_{p}\sqrt{L_{p}^{2} M^{2}-Q^{2}}
\end{equation}
where $L_{p}=\sqrt{G}$ is the Planck length in these units. The Hawking temperature and entropy of the black hole are functions of the location of the inner and outer horizons
\begin{equation}
\label{eq:TS}
T_{H}=\frac{r_{+}-r_{-}}{4\pi r_{+}^{2}},\quad S_{BH}=\frac{\pi r_{+}^{2}}{L_{p}^{2}}.
\end{equation}
Very near extremality ($M=Q/L_{p}$), the usual semi-classical analysis of the thermodynamic properties of the black hole breaks down \cite{PreskillSchwarz}. To see this, we define
\begin{equation}
E=M-\frac{Q}{L_{p}},
\end{equation}
the excitation energy above extremality. The horizons may then be written in terms of $E$ yielding
\begin{equation}
\label{eq:rpm}
r_{\pm}=Q L_{p}+E L_{p}^{2}\pm\sqrt{2 Q E L_{p}^{3}+E^{2}L_{p}^{4}}.
\end{equation}
For the near-extremal solution we have $E\ll 1$, hence substituting (\ref{eq:rpm}) into the expression for the Hawking temperature (\ref{eq:TS}) for small $E$ gives the energy-temperature relationship
\begin{equation}
\label{eq:excitationenergy}
E\sim 2\pi^{2} L_{p}Q^{3} T_{H}^{2}.
\end{equation}
When the energy of a typical quantum of Hawking radiation, which is of the order of the Hawking temperature $T_{H}$, is of the order of the excitation energy above extremality, the semi-classical analysis breaks down. This occurs at
\begin{equation}
T_{H}\sim E_{\textrm{gap}}\equiv \frac{1}{Q^{3}L_{p}},
\end{equation}
where in stringy black holes, the black holes have a mass gap and this energy represents the energy of the lowest-lying excitation \cite{MaldacenaSusskind}. 

In the extreme case, $E=0$, the horizons coincide $r_{h}=r_{\pm}=Q L_{p}$ and $T_{H}=0$. The line-element for extreme RN is
\begin{equation}
ds^{2}=-\frac{(r-r_{h})^{2}}{r^{2}}dt_{s}^{2}+\frac{r^{2}}{(r-r_{h})^{2}}dr^{2}+r^{2}d\Omega_{2}^{2}.
\end{equation}
The near-horizon limit is obtained by defining
\begin{equation}
\label{eq:lambda}
y=\frac{r-r_{h}}{L_{p}^{2}}
\end{equation}
and then considering the limit
\begin{equation}
L_{p}\rightarrow 0, \quad y, Q \textrm{ fixed}.
\end{equation}
The metric becomes
\begin{equation}
\frac{ds^{2}}{L_{p}^{2}}=-\frac{y^{2}}{Q^{2}}dt_{s}^{2}+\frac{Q^{2}}{y^{2}}dy^{2}+Q^{2}d\Omega_{2}^{2},
\end{equation}
which is the Bertotti-Robinson space-time with geometry $AdS_{2}\times\mathbb{S}^{2}$, or more accurately, $CAdS_{2}\times\mathbb{S}^{2}$, where $CAdS_{2}$ is the universal covering of anti-de Sitter space obtained by unwrapping the closed time-like curves. The form of this metric that we will find particularly useful is obtained by the transformation
\begin{equation}
t_{s}\pm\frac{Q^{2}}{y}=\tanh \Big(\frac{1}{2}\Big(t\pm\frac{1}{2}\ln\Big( \frac{\rho-1}{\rho+1}\Big)\Big)\Big)
\end{equation}
yielding
\begin{equation}
\label{eq:brmetric}
\frac{ds^{2}}{L_{p}^{2}Q^{2}}=-(\rho^{2}-1)dt^{2}+(\rho^{2}-1)^{-1}d\rho^{2}+d\Omega_{2}^{2}.
\end{equation}

We may also consider various inequivalent near-horizon limits for non-zero temperature. For example, in terms of the coordinate
\begin{equation}
U=\frac{r-r_{+}}{L_{p}^{2}},
\end{equation}
one can fix the geometry as the near-horizon limit is taken by keeping $T_{H}$ fixed, since the temperature is related to the periodicity of the time coordinate. We consider the limit
\begin{equation}
L_{p}\rightarrow 0\qquad Q, T_{H}\,\textrm{ fixed,}
\end{equation}
and the metric in this limit reads
\begin{align}
\label{eq:limit2}
\frac{ds^{2}}{Q^{2}L_{p}^{2}}=-\frac{U(U+4\pi Q^{2}T_{H})}{Q^{4}}dt_{s}^{2}+\frac{dU^{2}}{U(U+4\pi Q^{2}T_{H})}+d\Omega_{2}^{2}.
\end{align}
This may be recast in precisely the form of metric (\ref{eq:brmetric}) by the coordinate transformation
\begin{equation}
U=2\pi Q^{2}T_{H}(\rho-1),\quad t_{s}=\frac{1}{2\pi T_{H}}t.
\end{equation}

Yet another near-horizon limit of the Reissner Nordstrom black hole reduces to the metric of Eq.(\ref{eq:brmetric}), if we define
\begin{equation}
V=\frac{r-r_{+}}{Q^{2}L_{p}^{2}}
\end{equation}
and now hold $E$ and $T_{H}$ fixed, a near-horizon limit again emerges as $L_{p}\rightarrow 0$. In this case the charge $Q$ must diverge according to the scaling in Eq.(\ref{eq:excitationenergy}), i.e., we take the limit
\begin{equation}
L_{p}\rightarrow 0,\quad E, T_{H} \,\textrm{ fixed},\quad Q\sim L_{p}^{-1/2}\rightarrow\infty.
\end{equation}
The metric reduces to
\begin{align}
\Big(\frac{E_{\textrm{gap}}^{2/3}}{L_{p}^{4/3}}\Big)ds^{2}=-V(V+4\pi T_{H})dt_{s}^{2}+\frac{dV^{2}}{V(V+4\pi T_{H})}+d\Omega_{2}^{2}.
\end{align}
With the coordinate transformation
\begin{equation}
V=2\pi T_{H}(\rho-1),\quad t_{s}=\frac{1}{2\pi T_{H}}t,
\end{equation}
we again obtain the right-hand side of Eq.(\ref{eq:brmetric}).

\begin{figure}
\centering
\includegraphics[width=5cm]{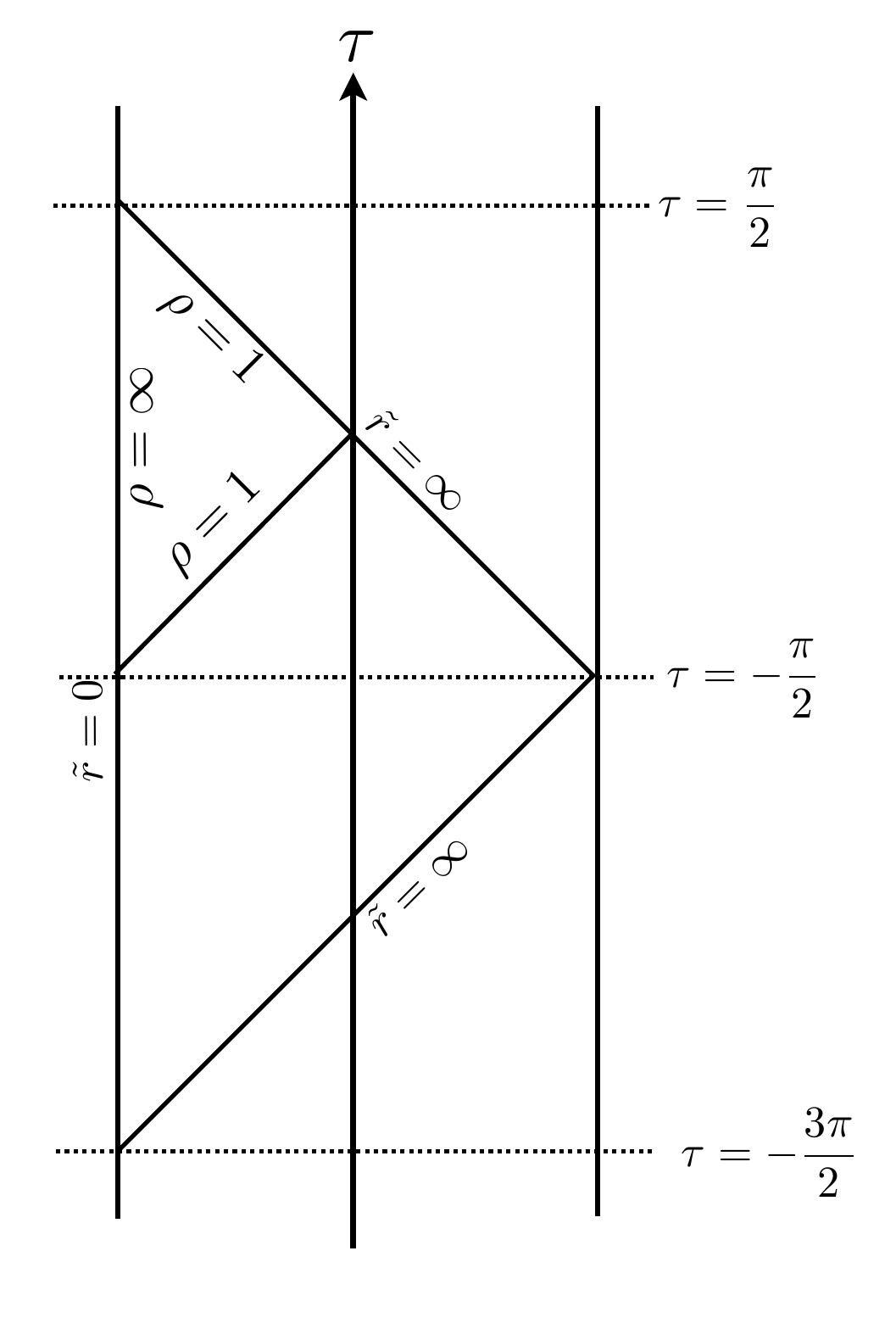}
\caption{{Penrose diagram showing the various coordinate patches on the Bertotti-Robinson space-time. It is clear from the diagram that the time $t$ only covers the part of the time-like boundary from $-\pi/2<\tau<\pi/2$ leading to a black hole interpretation in $(t, \rho)$ coordinates.} }
\label{fig:Penrose}
\end{figure}

So it is clear that the Bertotti-Robinson metric (\ref{eq:brmetric}) appears in a number of near-horizon limits of the Reissner-Nordstrom space-time. This form of the metric has a natural $AdS_{2}$ black hole interpretation even though the Bertotti-Robinson geometry is itself an electrovac solution to Einstein's field equation, not necessarily containing any horizons. This is due to the fact that we have inherited the time coordinate from the Reissner-Nordstrom solution which as we can see from the Penrose diagram in Fig.~\ref{fig:Penrose}, only covers part of the timelike boundary, which we call spatial infinity, of the covering space of $AdS_{2}$. The future ``black-hole'' horizon is then the boundary of the region from which nothing can escape to spatial infinity while the past horizon is the boundary of the region which cannot be accessed from spatial infinity.

This is analogous to the BTZ black hole \cite{BTZ} in three dimensions, where, despite the fact that all negative curvature spaces are locally equivalent to $AdS_{3}$, for certain global identifications there exists a black hole solution.

That we have a black-hole solution only for this preferred choice of time can be seen by transforming to global coordinates, for example, if we make the coordinate transformation
\begin{align}
U&=2\pi Q^{2}T_{H}(\sqrt{1+x^{2}}\cos\tau),\nonumber\\
 t_{s}&=\frac{1}{2\pi T_{H}}\tanh^{-1}\Big(\frac{\sqrt{1+x^{2}}}{x}\sin\tau\Big)
\end{align}
in the line-element (\ref{eq:limit2}), we obtain
\begin{equation}
\label{eq:globalmetric}
\frac{ds^{2}}{Q^{2}L_{p}^{2}}=-(1+x^{2})d\tau^{2}+(1+x^{2})^{-1}dx^{2}+d\Omega_{2}^{2}
\end{equation}
which covers all of $CAdS_{2}$ for $x\in(-\infty,\infty)$, $\tau\in (-\infty,\infty)$. On the other hand, only the part of the time-like boundary from $-\pi/2<\tau<\pi/2$ is covered by the coordinates $(t, \rho)$ which is easily seen from their coordinate relations
\begin{align}
\label{eq:globalcoords}
x&=(\rho^{2}-1)^{1/2}\cosh t,\nonumber\\
\tau&=\sin^{-1}\Big[\frac{(\rho^{2}-1)^{1/2}\sinh t}{(1+(\rho^{2}-1)\cosh^{2} t)^{1/2}}\Big].
\end{align}

Each new coordinate system defines a new vacuum, by choosing normalized modes to be positive frequency with respect to a particular time coordinate. For example, taking modes to be positive frequency with respect to $t$ in the metric (\ref{eq:brmetric}) defines the Boulware vacuum while taking modes to be positive frequency with respect to the global coordinates (\ref{eq:globalmetric}) defines the Global vacuum. Other common choices are the Poincar\'e vacuum and the Schwarzschild vacuum. The Poincar\'e vacuum is defined by choosing positive frequency modes with respect to Poincar\'e coordinates for which the metric is conformally flat. Making the coordinate transformations
\begin{align}
\label{eq:poincarecoords}
\tilde{t}=\frac{(\rho^{2}-1)^{1/2}\sinh t}{\rho+(\rho^{2}-1)^{1/2}\cosh t},\quad\tilde{r}=\frac{1}{\rho+(\rho^{2}-1)^{1/2}\cosh t},
\end{align}
in Eq.(\ref{eq:brmetric}), we arrive at the Poincar\'e form of the metric,
\begin{equation}
\label{eq:conformalflat}
\frac{ds^{2}}{Q^{2}L_{p}^{2}}=\tilde{r}^{-2}(-d\tilde{t}^{2}+d\tilde{r}^{2}+\tilde{r}^{2}d\Omega_{2}^{2}).
\end{equation}
Finally, the Schwarzschild coordinates are defined by
\begin{equation}
\label{eq:schwtransform}
\rho^{*}=\frac{1}{4\pi T_{H}}\ln \frac{\rho-1}{\rho+1},
\end{equation}
which yields
\begin{equation}
\frac{ds^{2}}{Q^{2}L_{p}^{2}}=\Big(\frac{2\pi T_{H}}{\sinh(2\pi T_{H}\rho^{*})}\Big)^{2}(-dt^{2}+d\rho^{*}{}^{2})+d\Omega_{2}^{2}.
\end{equation}
Since the transformation (\ref{eq:schwtransform}) is independent of the time coordinates, the Schwarzschild vacuum is equivalent to the Boulware vacuum.

\section{Vacuum States and Propagators on Bertotti-Robinson}
\label{sec:propagators}
In this section, we derive the Feynman Green's function for massless scalars for the various quantum states. Of the vacua already mentioned, only two are distinct, the Boulware and Hartle-Hawking states \cite{StromingerAdSBH}. We will also consider the case of a field at an arbitrary temperature, not necessarily equal to the Hawking temperature. For the remainder of this paper, we set $L_{p}^{2} Q^{2}=1$ which may be restored by dimensional analysis.

The physical quantity of interest in the semi-classical theory is the renormalized expectation value of the stress-energy tensor in a unit-norm quantum state $\big|A\big>$. This is closely related to the Feynman Green's function which is defined by
\begin{equation}
G_{A}(x,x')=i\big<A\big|T\{\hat{\varphi}(x), \hat{\varphi}(x')\}\big|A\big>
\end{equation}
where $T$ denotes the time-ordered product of the quantum field operators such that
\begin{align}
T\{\hat{\varphi}(t,\textbf{x}), \hat{\varphi}(t',\textbf{x}')\}=\begin{cases}
\hat{\varphi}(t,\textbf{x}) \hat{\varphi}(t',\textbf{x}')\quad &\textrm{if}\,\,\,t>t'\\
\hat{\varphi}(t',\textbf{x}') \hat{\varphi}(t,\textbf{x})\quad &\textrm{if}\,\,\,t'>t.
\end{cases}
\end{align}
The field operator $\hat{\varphi}(x)$ is expanded in terms of normalized mode-functions which satisfy the homogeneous wave equation whereby the choice of boundary conditions on these mode-functions determine the quantum state. It is straight-forward to show that the canonical commutation relations satisfied by $\hat{\varphi}(x)$ imply that the Feynman Green's function satisfies the inhomogeneous wave equation \cite{Fulling}
\begin{equation}
\Box G_{A}(x,x')=-g^{-1/2}\delta(x-x')
\end{equation}
where $\Box=g^{\mu\nu}\nabla_{\mu}\nabla_{\nu}$ is the d'Alembertian operator and $g=|\det\,g_{\mu\nu}|$ is the metric determinant. Since the Bertotti-Robinson metric has vanishing scalar curvature, the Green's function does not depend on the field's coupling to the curvature, though the stress-energy tensor will. We will find it convenient to express the Feynman Green's function in terms of the Wightman function and its complex conjugate,
\begin{equation}
\label{eq:gfeynman}
G_{A}(x,x')=i \theta (\Delta t)G_{A}^{+}(x,x')+i \theta (-\Delta t)G_{A}^{-}(x,x')
\end{equation}
where $\Delta t=t-t'$ and
\begin{equation}
\label{eq:wightman}
G_{A}^{+}(x,x')=\left<A| \hat{\varphi}(x)\hat{\varphi}(x')|A\right>
\end{equation}
is the Wightman function with $G_{A}^{-}(x,x')$ as its complex conjugate.

We now consider the pure, zero-temperature vacuum states and the mixed thermal states separately.

\subsection{Boulware Vacuum}
The Boulware vacuum, which we denote by $\big|B\big>$, is defined by requiring that the normal modes are positive frequency with respect to the time-like Killing vector $\partial/\partial t$.

The homogeneous wave equation is
\begin{align}
&\Big[-\frac{1}{(\rho^{2}-1)}\frac{\partial^{2}}{\partial t^{2}}+\frac{\partial}{\partial\rho}\Big((\rho^{2}-1)\frac{\partial}{\partial \rho}\Big)\nonumber\\
&\qquad+\frac{1}{\sin\theta}\frac{\partial}{\partial\theta}\Big(\sin\theta\frac{\partial}{\partial\theta}\Big)
+\frac{1}{\sin^{2}\theta}\frac{\partial^{2}}{\partial\phi^{2}}\Big]\varphi(x)=0
\end{align}
A complete set of mode functions may be obtained by a separation of variables:
\begin{align}
\label{eq:modesoln}
u_{\omega l m}(x)=N_{\omega}e^{-i\omega t}Y_{lm}(\theta, \phi)R_{\omega l}(\rho)
\end{align}
where $Y_{lm}(\theta,\phi)$ are the spherical harmonics normalized such that
\begin{equation}
\sum_{m=-l}^{l}Y_{lm}(\theta,\phi)Y_{lm}^{*}(\theta,\phi)=\frac{2 l+1}{4\pi}.
\end{equation}
and $R_{\omega l}(r)$ satisfies
\begin{equation}
\label{eq:radialeq}
\Big(\frac{d}{d\rho}\Big((\rho^{2}-1)\frac{d}{d\rho}\Big)+\frac{\omega^{2}}{(\rho^{2}-1)}-l(l+1)\Big)R_{\omega l}(\rho)=0.
\end{equation}
Solutions of this equation are the associated Legendre functions of pure imaginary order, $P_{l}^{\pm i\omega}(\rho)$ and $Q_{l}^{\pm i\omega}(\rho)$, where the specific combination of these solutions is determined by the boundary conditions on the field. Much like in AdS space-time, the Bertotti-Robinson space-time is not globally hyperbolic and it possesses a time-like boundary at spatial infinity through which information can propagate. We impose vanishing boundary conditions at spatial infinity since this time-like surface is an infinite proper distance from any finite radius. This condition rules out the associated Legendre functions $P_{l}^{\pm i\omega}(\rho)$ in our mode-function expansion of the field operator since these functions diverge as $\rho\rightarrow \infty$. Choosing $R_{\omega l}(\rho)=Q_{l}^{i\omega}(\rho)$, then the normalization condition
\begin{equation}
\left<u_{\omega l m}, u_{\omega' l' m'}\right>=\delta(\omega-\omega')\delta_{ll'}\delta_{mm'},
\end{equation}
fixes the normalization constant to be
\begin{equation}
\label{eq:normconstant}
N_{\omega}=\frac{1}{\pi}e^{\omega \pi} \sqrt{\sinh( \omega\pi)},
\end{equation}
where
\begin{equation}
\left<\varphi_{1}, \varphi_{2}\right>=i\int_{\Sigma}(\varphi_{1}^{*}\partial_{\mu}\varphi_{2}-\varphi_{2}\partial_{\mu}\varphi_{1}^{*}) n^{\mu}d\Sigma
\end{equation}
is the Klein-Gordon inner product.

The field operator expanded in terms of normalized mode functions is therefore
\begin{equation}
\hat{\varphi}(x)=\int_{0}^{\infty}d\omega \sum_{l=0}^{\infty}\sum_{m=-l}^{l}u_{\omega l m}(x)\hat{a}_{\omega l m}+u_{\omega l m}^{*}(x)\hat{a}_{\omega l m}^{\dagger}
\end{equation}
where $u_{\omega l m}$ is given by Eqs. (\ref{eq:modesoln}) and (\ref{eq:normconstant}). It is straightforward to show that the commutation relations satisfied by the field operator and its conjugate momentum are equivalent to the commutation relations
\begin{align}
[\hat{a}_{\omega l m}, \hat{a}_{\omega' l' m'}^{\dagger}]&=\delta(\omega-\omega')\delta_{ll'}\delta_{m m'},\nonumber\\
[\hat{a}_{\omega l m}, \hat{a}_{\omega' l' m'}]&=[\hat{a}_{\omega l m}^{\dagger}, \hat{a}_{\omega' l' m'}^{\dagger}]=0.
\end{align}

To obtain the Feynman Green's function, we compute the Wightman function defined by Eq.(\ref{eq:wightman}). In terms of the mode functions, this is given by
\begin{align}
G^{+}_{B}(x,x')&=\int_{0}^{\infty}d\omega \sum_{l=0}^{\infty}\sum_{m=-l}^{l}u_{\omega l m}(x) u_{\omega l m}^{*}(x')\nonumber\\
&=\frac{1}{4\pi^{3}}\int_{0}^{\infty}d\omega\> e^{-i\omega \Delta t}e^{2\omega\pi}\sinh(\pi \omega)\nonumber\\
&\quad \times \sum_{l=0}^{\infty}(2l+1)P_{l}(\cos\gamma)Q^{i\omega}_{l}(\rho)[Q^{i\omega}_{l}(\rho')]^{*},
\end{align}
where $\cos\gamma=\cos\theta\cos\theta'+\sin\theta\sin\theta'\cos(\phi-\phi')$.
Using standard results relating associated Legendre functions \cite{gradriz}, the Wightman function may be recast in the following form,
\begin{align}
G^{+}_{B}(x,x')=\frac{i}{8\pi^{2}}\int_{0}^{\infty}d\omega \> e^{-i\omega \Delta t}\sum_{l=0}^{\infty}(2l+1)P_{l}(\cos\gamma)\nonumber\\
\times\Big[e^{\omega\pi}P^{-i\omega}_{l}(\rho')Q^{i\omega}_{l}(\rho)-e^{-\omega\pi}P^{i\omega}_{l}(\rho')Q^{-i\omega}_{l}(\rho)\Big]
\end{align}
The particular combination of associated Legendre functions appearing in square brackets above is invariant under the interchange of the radial coordinates and hence we have
\begin{align}
\label{eq:gwightman}
G^{+}_{B}(x,x')=\frac{i}{8\pi^{2}}\int_{0}^{\infty}d\omega \>e^{-i\omega \Delta t}\sum_{l=0}^{\infty}(2l+1)P_{l}(\cos\gamma)\nonumber\\
\times\Big[e^{\omega\pi}P^{-i\omega}_{l}(\rho_{<})Q^{i\omega}_{l}(\rho_{>})-e^{-\omega\pi}P^{i\omega}_{l}(\rho_{<})Q^{-i\omega}_{l}(\rho_{>})\Big]
\end{align}
where $\rho_{<}=\min\{\rho,\rho'\}$ and $\rho_{>}=\max\{\rho,\rho'\}$. Performing the $l$ sum in this expression is crucial if we are to obtain a closed-form representation of the Green's function for the Boulware vacuum. This is a non-standard summation formula and to the best of our knowledge has never been published, except for the case of associated Legendre functions of real integer order \cite{Candelas:1980zt}. Rather than include the derivation here, we defer it to the Appendix. The result is
\begin{align}
\label{eq:legendresum}
\sum_{l=0}^{\infty}(2 l+1) P_{l}(\cos\gamma) e^{-i \mu \pi} P_{l}^{-\mu}(\rho_{<})Q_{l}^{\mu}(\rho_{>})=\frac{e^{-\mu \eta}}{R^{1/2}},
\end{align}
where $\mu$ is an arbitrary complex constant and
\begin{align}
\label{eq:defRChi}
\cosh\eta&=\frac{\rho\rho'-\cos\gamma}{(\rho^{2}-1)^{1/2}(\rho'^{2}-1)^{1/2}},\nonumber\\
R&=\rho^{2}+\rho'^{2}-2\rho\rho'\cos\gamma-\sin^{2}\gamma\nonumber\\
&=(\rho^{2}-1)(\rho'^{2}-1)\sinh^{2}\eta.
\end{align}
Using the result (\ref{eq:legendresum}) for $\mu=\pm i\omega$ in Eq.(\ref{eq:gwightman}), we obtain
\begin{align}
G^{+}_{B}(x,x')=\frac{1}{4\pi^{2}R^{1/2}}\int_{0}^{\infty}d\omega\>e^{-i\omega \Delta t}\sin(\omega\, \eta).
\end{align}

The $\omega$ integral is not immediately convergent which is a consequence of the fact that the Green's function is a distributional quantity, being the solution of a wave equation with a delta distribution source. We can obtain a closed-form representation by the usual `$i\epsilon$' prescription \cite{BD} by adding a small negative imaginary part to the time coordinate so that the integral converges and then taking the $\epsilon\rightarrow 0^{+}$,
\begin{align}
G^{+}_{B}(x,x')=\frac{1}{4\pi^{2}R^{1/2}}\lim_{\epsilon\rightarrow 0^{+}}\int_{0}^{\infty}d\omega\>e^{-i\omega (\Delta t-i\epsilon)}\sin(\omega \,\eta).
\end{align}
The integral is now trivial, yielding
\begin{align}
G^{+}_{B}(x,x')=\frac{1}{4\pi^{2}R^{1/2}}\lim_{\epsilon\rightarrow 0^{+}}\frac{\eta}{(-(\Delta t-i\epsilon)^{2}+\eta^{2})}.
\end{align}
From the definition $(\ref{eq:gfeynman})$, the Feynman Green's function for the Boulware vacuum is
\begin{align}
G_{B}(x,x')=\frac{i}{4\pi^{2}R^{1/2}}\lim_{\epsilon\rightarrow 0^{+}}\frac{\eta}{(-(|\Delta t|-i\epsilon)^{2}+\eta^{2})}.
\end{align}
For small $\epsilon$, we can rewrite this as
\begin{align}
G_{B}(x,x')=\frac{i}{4\pi^{2}R^{1/2}}\lim_{\epsilon\rightarrow 0^{+}}\frac{\eta}{(-\Delta t^{2}+\eta^{2}+i\epsilon)},
\end{align}
where we have absorbed the $2|\Delta t|$ into the definition of $\epsilon$ and we have ignored $O(\epsilon^{2})$ terms. We now make use of the distributional identity
\begin{equation}
\label{eq:distidentity}
\frac{1}{z^{2}+i\epsilon}=\mathcal{P}\frac{1}{z^{2}}-\pi i\delta(z^{2}),\quad\epsilon\rightarrow 0^{+},
\end{equation}
where $\mathcal{P}$ denotes the principal part, to obtain
\begin{align}
\label{eq:gboulware}
G_{B}(x,x')&=\frac{i}{4\pi^{2}R^{1/2}}\mathcal{P}\frac{\eta}{(-\Delta t^{2}+\eta^{2})}\nonumber\\
&\qquad+\frac{\eta}{4\pi R^{1/2}}\,\delta(-\Delta t^{2}+\eta^{2}).
\end{align}

\subsection{Thermal States}
The Boulware Green's function is a zero temperature propagator since it is the expectation value of a product of field operators in a pure state, namely the Boulware vacuum. We can also consider mixed thermal states where the field is at some arbitrary temperature. Moreover, if we consider the Bertotti-Robinson geometry arising as the near-extremal Reissner-Nordstr\"om black hole, then we can define the Schwarzschild analog of the Hartle-Hawking ``vacuum'' \cite{HartleHawking} which corresponds to the field in thermal equilibrium with the black hole horizon and its defining feature is regularity on the past and future horizon.

We will make use of the following relation between the Feynman Green's function, the symmetric Green's function, $\bar{G}(x,x')$ which is the average of the advanced and retarded Green's functions and the Hadamard two point function, $G^{(1)}(x,x')$,
\begin{align}
G_{A}(x,x')=\bar{G}(x,x')+\tfrac{1}{2}i G_{A}^{(1)}(x,x').
\end{align}
The retarded and advanced Green's function, and hence $\bar{G}(x,x')$, are expectation values of ``c-number'' operators (multiples of the identity) and hence do not depend on the vacuum state. On the other hand, the definition of the Hadamard function hinges on the decomposition of mode solutions into positive and negative frequency parts and therefore depends on the vacuum. The thermal Hadamard function is given as an infinite imaginary-time image sum of the corresponding zero-temperature Hadamard function \cite{BD},
\begin{equation}
G^{(1)}_{\beta}(x,x')=\sum_{k=-\infty}^{\infty}G^{(1)}_{B}(t+i k \beta,\textbf{x}; t', \textbf{x}'),
\end{equation}
where $\beta=1/T$ is the inverse temperature of the field. It follows that the Feynman Green's function for a mixed thermal state is
\begin{align}
G_{\beta}(x,x')&=\frac{i}{4\pi^{2}R^{1/2}}\mathcal{P}\sum_{k=-\infty}^{\infty}\frac{\eta}{(-(\Delta t+i k  \beta)^{2}+\eta^{2})}\nonumber\\
&\qquad +\frac{\eta}{4\pi R^{1/2}}\,\delta(-\Delta t^{2}+\eta^{2}).
\end{align}
The $k$ sum may be performed yielding a closed-form propagator for a massless field at temperature $T$,
\begin{align}
G_{\beta}(x,x')&=\frac{i T}{4\pi R^{1/2}}\mathcal{P}\frac{\sinh(2\pi T \eta)}{(\cosh(2\pi T \eta)-\cosh(2\pi T \Delta t))}\nonumber\\
&\qquad +\frac{\eta}{4\pi R^{1/2}}\,\delta(-\Delta t^{2}+\eta^{2}).
\end{align}
To the best of our knowledge, this result has not previously been given in the literature.

For the near-extremal Reissner-Nordstr\"om limit, we can define the Hartle-Hawking state whereby the field is in thermal equilibrium with the black hole at the Hawking temperature $T_{H}=\kappa/2\pi$, where $\kappa$ is the surface gravity given by
\begin{equation}
\kappa^{2}=\frac{\nabla^{\lambda}(k^{\mu}k_{\mu})\nabla_{\lambda}(k^{\nu}k_{\nu})}{4 k^{\sigma}k_{\sigma}}\Bigg|_{\rho\rightarrow 1}=1,
\end{equation}
where $k^{\mu}=(1,0,0,0)$ is a Killing vector normal to the horizon. The Green's function now simplifies considerably,
\begin{align}
&G_{H}(x,x')=\nonumber\\
&\qquad\frac{i}{8\pi^{2}(\rho^{2}-1)^{1/2}(\rho'^{2}-1)^{1/2}}\mathcal{P}\frac{1}{(\cosh\eta-\cosh\Delta t)}\nonumber\\
&\qquad\qquad +\frac{\eta}{4\pi R^{1/2}}\,\delta(-\Delta t^{2}+\eta^{2}),
\end{align}
where $\eta$ and $R$ are given by Eq.(\ref{eq:defRChi}). This can be recast into the succinct form
\begin{align}
\label{eq:ghartlehawking}
G_{H}(x,x')=\frac{i}{8\pi^{2}}\mathcal{P}\frac{1}{\cosh\lambda-\cos\gamma}+\frac{1}{8\pi}\delta(\cosh\lambda-\cos\gamma),
\end{align}
where $\lambda$ is the geodesic distance on $AdS_{2}$ satisfying
\begin{equation}
\label{eq:lambda}
\cosh\lambda=\rho\rho'-(\rho^{2}-1)^{1/2}(\rho'^{2}-1)^{1/2}\cosh\Delta t
\end{equation}
and $\gamma$ is the geodesic distance on the two-sphere satisfying
\begin{equation}
\label{eq:gamma}
\cos\gamma=\cos\theta\cos\theta'+\sin\theta\sin\theta'\cos(\phi-\phi').
\end{equation}

One can verify that the Hartle-Hawking state is equivalent to the Global and Poincar\'e states by explicitly constructing the Green's functions. For example, the normalized modes in Poincar\'e coordinates are
\begin{equation}
u_{\omega l m}=\frac{1}{\sqrt{2}}e^{-i\omega \tilde{t}} Y_{lm}(\theta, \phi) \tilde{r}^{1/2}J_{l+1/2}(\omega \tilde{r}).
\end{equation}
Following a similar procedure to the previous section, we obtain the following mode-sum expression for the Wightman function
\begin{align}
G^{+}_{P}(x,x')=\frac{1}{8\pi}\int_{0}^{\infty}e^{-i\omega\, \Delta \tilde{t}}d\omega \sum_{l=0}^{\infty}(2l+1)P_{l}(\cos\gamma)\nonumber\\
\times \tilde{r}^{1/2}\tilde{r}'^{1/2}J_{l+1/2}(\omega \tilde{r})J_{l+1/2}(\omega \tilde{r}').
\end{align}
The $l$ sum may be performed using standard summation formulae for Bessel functions \cite{gradriz} yielding
\begin{equation}
G^{+}_{P}(x,x')=\frac{\tilde{r} \tilde{r}'}{4\pi^{2}\zeta}\int_{0}^{\infty}e^{-i\omega\,\Delta\tilde{t}}\sin\omega\zeta\,d\omega,
\end{equation}
where
\begin{equation}
\zeta=\sqrt{\tilde{r}^{2}+\tilde{r}'^{2}-2 \tilde{r} \tilde{r}'\cos\gamma}.
\end{equation}
As before, the integral here does not converge and we follow the distributional treatment of the previous subsection to obtain the following representation of the Feynman Green's function in the Poincar\'e vacuum,
\begin{align}
G_{P}(x,x')=\frac{i}{4\pi^{2}}\lim_{\epsilon\rightarrow 0^{+}}\frac{\tilde{r} \tilde{r}'}{(-\Delta\tilde{t}^{2}+\zeta^{2}+i\epsilon)}.
\end{align}
Employing the identity (\ref{eq:distidentity}) again gives
\begin{equation}
\label{eq:greensfnpoincare}
G_{P}(x,x')=\frac{i}{8\pi^{2}}\mathcal{P}\frac{1}{\cosh\lambda-\cos\gamma}+\frac{1}{8\pi}\delta(\cosh\lambda-\cos\gamma)
\end{equation}
where, in Poincar\'e coordinates, the geodesic distance on $AdS_{2}$ is defined by
\begin{equation}
\cosh\lambda=\frac{-\Delta\tilde{t}^{2}+\tilde{r}^{2}+\tilde{r}'^{2}}{2\tilde{r}\tilde{r}'}.
\end{equation}
This form of the Green's function has been previously given by Kofman and Sahni \cite{KofmanSahni1}. It is clearly identical to the Hartle-Hawking propagator (\ref{eq:ghartlehawking}). A similar calculation reveals that it is also identical to the Green's function in the Global vacuum.

\section{The renormalized stress-energy tensor}
\label{sec:rset}
The stress-energy tensor operator ought to provide an absolute measure of the energy-momentum density for the fields in the state $|A\rangle$ as well as governing the back-reaction on the classical gravitational background via the semi-classical Einstein equations
\begin{equation}
\label{eq:semiclassicaleqns}
R^{ab}-\frac{1}{2}g^{ab}R=8\pi \langle A|\hat{T}^{ab} |A\rangle,
\end{equation}
where classically the stress energy tensor for a massless field $\varphi$ in a space-time with vanishing scalar curvature is given by
\begin{align}
\label{eq:stresstensorclassical}
T^{ab}=(1-2\xi)\varphi^{;a}\varphi^{;b}+(2\xi-\tfrac{1}{2})g^{ab}\varphi_{;c}\varphi^{;c}\nonumber\\
-2\xi\,\varphi\,\varphi^{;ab}+2\xi\,g^{ab}\varphi\,\Box\varphi+\xi\,R^{ab}\varphi^{2},
\end{align}
where $\xi$ is the field's coupling to the scalar curvature in the action.
The right-hand side of Eq.(\ref{eq:semiclassicaleqns}) is clearly divergent reflecting the fact that it involves products of operator-valued distributions evaluated at the same space-time point and therefore requires renormalization. It proves useful to write it as a coincidence limit of a `point-split' stress-energy tensor operator
\begin{equation}
\label{eq:stressenergyunren}
\langle A|\hat{T}^{ab}|A\rangle\equiv -i[\hat{\tau}^{ab} G_{A}(x,x')]=-i\lim_{x\rightarrow x'}\hat{\tau}^{ab}G_{A}(x,x').
\end{equation}
where $G_{A}(x,x')$ is the Feynman Green's function for the quantum state $\left|A\right>$ and $\hat{\tau}^{ab}$ is a differential operator defined so that $\hat{\tau}^{ab}\varphi(x)\varphi(x')$ yields (\ref{eq:stresstensorclassical}) in the coincidence limit. For example, we take 
\begin{align}
\hat{\tau}^{ab}=(1-2\xi)g_{b'}{}^{b}\nabla^{a}\nabla^{b'}+(2\xi-\tfrac{1}{2})g^{ab}g_{c'}{}^{c}\nabla^{c}\nabla^{c'}\nonumber\\
-2\xi\nabla^{a}\nabla^{b}+2\xi g^{ab}\nabla_{c}\nabla^{c}+\xi\,R^{ab},
\end{align}
where $g_{a'}{}^{b}$ denotes the bivector of parallel transport satisfying the differential equation $\sigma^{;c}g_{a'}{}^{b}{}_{;c}=0$ subject to the boundary condition $[g_{a'}{}^{b}]=\delta_{a}{}^{b}$, so that $g_{a'}{}^{b} u^{a'}$ is the result of parallel transporting $u^{a'}$ along the geodesic from $x'$ to $x$. 

We shall employ the Hadamard renormalization prescription \cite{BrownOttewill1983, BrownOttewill1986} which relies on the Hadamard representation \cite{Hadamard} of the Feynman Green's function corresponding to a unit-norm state $\big| A\big>$,
\begin{align}
\label{eq:ghadamard}
G_{A}(x,x')&=\frac{i}{8\pi^{2}}\Big(\frac{\Delta^{1/2}(x,x')}{\sigma(x,x')+i\epsilon}+V(x,x')\ln(\sigma(x,x')+i\epsilon)\nonumber\\
&\qquad+W_{A}(x,x')\Big).
\end{align}
where $\sigma(x,x')$ is Synge's world function, $\Delta(x,x')$ is the Van Vleck-Morette determinant and $V(x,x')$ and $W_{A}(x,x')$ are symmetric, regular biscalars. Global information such as the boundary conditions and information about the quantum state is encoded in $W_{A}$. We have added the usual $i\epsilon$ prescription to indicate that $G_{A}$ is really the boundary value of a function which is analytic in the lower-half $\sigma$ plane. The geometrical nature of the divergences inflicting the expectation value of the stress-energy tensor operator is explicit in the Hadamard representation of the Feynman Green's function, where $\sigma$ vanishes in the coincidence limit.

Substituting the Hadamard representation of the Green's function into the wave equation and equating logarithmic and power terms separately yields
\begin{align}
\label{eq:veqn}
\Box V&=0,\\
\label{eq:weqn}
\sigma\Box W_{\!A}&=-2 V-2\sigma^{;a}(V_{;a}-V \Delta^{-1/2}(\Delta^{1/2})_{;a})\nonumber\\
&\qquad-\Box(\Delta^{1/2}).
\end{align}
Moreover, $V$ and $W_{A}$ can be expanded in powers of $\sigma$,
\begin{align}
V(x,x')&=\sum_{n=0}^{\infty}V_{n}(x,x')\sigma^{n},\nonumber\\
W_{\!A}(x,x')&=\sum_{n=0}^{\infty}W_{\!A\,n}(x,x')\sigma^{n},
\end{align}
which when substituted into Eqs.(\ref{eq:veqn})-(\ref{eq:weqn}) gives a set of recursion relations that completely determine the $V_{n}$ (for $n\ge0$) and also determines the $W_{\!A\,n}$ (for $n\ge1$) once $W_{\!A\,0}$ is given.

The spirit of the point-splitting renormalization scheme is to subtract from the Green's function an appropriate parametrix that results in a finite quantity in the coincidence limit. The obvious candidate parametrix is the direct part of the Hadamard representation of the Green's function, which upon subtraction gives
\begin{align}
\label{eq:Tfinite}
\langle A|\hat{T}^{ab}|A\rangle_{\textrm{finite}}\equiv \frac{1}{8\pi^{2}}[\hat{\tau}^{ab}W_{\!A}(x,x')].
\end{align}
Although this is finite, in general it is not conserved, a property satisfied by the classical stress-energy tensor and required for consistency of the semi-classical Einstein equations. The classical conservation equations are
\begin{align}
T^{ab}{}_{;b}=\varphi^{;a}\, \Box\varphi=0,
\end{align}
which implies the following point-split quantity
\begin{equation}
[\hat{\tau}^{ab}{}_{;b} W_{\!A}(x,x')]=[g^{a a'}\nabla_{a'}\,\Box W_{\!A}(x,x')].
\end{equation}
The recursion relations for $V_{n}$ may be used in Eq.(\ref{eq:weqn}) to show that
\begin{align}
\label{eq:boxw}
\Box W_{\!A}(x,x')=-6 v_{1}(x)+2 v_{1}(x)_{;a}\sigma^{;a}+\mathcal{O}(\sigma),
\end{align}
where we have used the covariant Taylor expansion $V_{1}(x,x')=v_{1}(x)-\tfrac{1}{2}v_{1}(x)_{;a}\sigma^{;a}+\cdot\cdot\cdot$ where $v_{1}(x)=[V_{1}(x,x')]$. It follows that
\begin{align}
[\hat{\tau}^{ab}{}_{;b} W_{A}(x,x')]=-2 v_{1}(x)^{;a},
\end{align}
so that the conserved, renormalized stress-energy tensor operator is
\begin{align}
\label{eq:Tren}
\langle A| \hat{T}^{ab} |A\rangle_{\textrm{ren}}\equiv \frac{1}{8\pi^{2}}\Big([\hat{\tau}^{ab} W_{A}(x,x')]+2 v_{1}(x) g^{ab}\Big).
\end{align}
The biscalar $W_{A}(x,x')$ may be covariantly Taylor expanded as 
\begin{align}
\label{eq:wTaylor}
W_{\!A}(x,x')&=w_{A}(x)-\tfrac{1}{2}w_{A}(x)_{;a}\sigma^{;a}\nonumber\\
&\qquad+\tfrac{1}{2}\varpi_{A\,ab}(x)\sigma^{;a}\sigma^{;b}
+\mathcal{O}(\sigma^{3/2}),
\end{align}
where $w_{A}(x)=[W_{\!A}(x,x')]$ and $\varpi_{A\,ab}(x)$ is a second-rank tensor at $x$. The odd-order coefficients may be determined recursively by using the the fact that $W_{\!A}(x,x')$ is symmetric in its arguments \cite{BrownOttewill1986}. Note that, as a consequence of the defining equation for $\sigma$,
$2\sigma=\sigma^{;a}\sigma_{;a}$,
we must have
$\mathcal{O}(\sigma^{;a})\sim \mathcal{O}(\sigma^{1/2})$.
Substituting the Taylor expansion into Eq.(\ref{eq:Tren}) and taking coincidence limits \cite{Christensen} yields
\begin{align}
\label{eq:Trenw}
\langle A| \hat{T}^{a}{}_{b}|A\rangle_{\textrm{ren}}=\frac{1}{8\pi^{2}}\Big(\tfrac{1}{2}(1-2\xi)w_{A}{}^{;a}{}_{b}+\tfrac{1}{2}(2\xi-\tfrac{1}{2})\delta^{a}{}_{b}\Box w_{A}\nonumber\\
-\varpi_{A}{}^{a}{}_{b}+\tfrac{1}{2}\delta^{a}{}_{b}\varpi_{A}{}^{c}{}_{c}+\xi R^{a}{}_{b} w_{A}+2 \delta^{a}{}_{b}v_{1}\Big).
\end{align}
Similarly, substituting the Taylor expansion into Eq.(\ref{eq:boxw}) and taking coincidence limits gives the constraint
\begin{equation}
\label{eq:wconstraint}
\varpi_{A}(x)^{a}{}_{a}=-6 v_{1}(x),
\end{equation}
which may be employed in obtaining the trace of the stress-energy tensor,
\begin{equation}
\langle A| \hat{T}^{a}{}_{a}|A\rangle_{\textrm{ren}}=\frac{1}{8\pi^{2}}\Big(2 v_{1}(x)+3(\xi-\tfrac{1}{6})\Box w_{A}(x)\Big).
\end{equation}
Hence, we reproduce the standard trace anomaly for a conformally coupled scalar field. For massless scalar fields, we have
\begin{align}
v_{1}(x)&=\tfrac{1}{720}R_{abcd}R^{abcd}-\tfrac{1}{720}R_{ab}R^{ab}+\tfrac{1}{8}(\xi-\tfrac{1}{6})^2 R^2\nonumber\\
&\qquad-\tfrac{1}{24}(\xi-\tfrac{1}{5})\square R.
\end{align}
In Bertotti-Robinson space-time in our conventions
\begin{align}
R_{abcd}R^{abcd}=8, \quad R_{ab}R^{ab}=4,\quad R=0
\end{align}
so
\begin{align}
v_1 = \tfrac{1}{180}.
\end{align}
Note that, in this particular case, the finite stress-energy tensor in Eq.(\ref{eq:Tfinite}) is already conserved and one need not add the geometrical term involving $v_{1}(x)$ in Eq.(\ref{eq:Tren}), though its inclusion is necessary if we are to obtain the standard trace anomaly so we prefer the latter definition.

To obtain explicit expressions for the components of the renormalized stress-energy tensor in Eq.(\ref{eq:Trenw}), we require the global terms $w_{A}(x)$ and $\varpi_{A}(x)_{ab}$ which cannot be obtained directly from the quasi-local Hadamard expansion but by subtracting the Hadamard parametrix from the globally defined closed-form expressions derived in Sec.~\ref{sec:propagators}. This is a particularly neat method in the present case where $\Delta^{1/2}$ and $\sigma$ can be calculated in closed form and $V\equiv0$ for conformally flat space-times. 
Note that Bertotti-Robinson space-time falls into the class of space-times where $V(x,x')=0$ but $v_{1}(x)\neq 0$ so that the choice $W(x,x')=0$ does not define a Green's function.

For direct product space-times, $\sigma$ factorizes as a sum of the square of the geodesic distances on the two parts of the space-time. For the Bertotti-Robinson space-time, we have
\begin{equation}
\sigma=\sigma_{AdS}+\sigma_{S^{2}}=\tfrac{1}{2}\lambda^{2}+\tfrac{1}{2}\gamma^{2},
\end{equation}
where $\lambda$ is the geodesic distance on the $AdS_{2}$ part of the metric which in $(t,\rho)$ coordinates is given by Eq.(\ref{eq:lambda}) while $\gamma$ is the geodesic distance on $\mathbb{S}^{2}$ given by Eq.(\ref{eq:gamma}).
The Van Vleck-Morette determinant correspondingly factorizes as the product of determinants on the two parts of the space-time,
\begin{equation}
\Delta=\Delta_{AdS}\Delta_{S^{2}},
\end{equation}
where
\begin{align}
\Delta_{AdS}=\frac{\lambda}{\sinh \lambda},\quad \Delta_{S^{2}}=\frac{\gamma}{\sin\gamma}.
\end{align}
Hence the direct part of the Hadamard form is
\begin{equation}
\label{eq:Gdiv}
G_{\textrm{div}}(x,x')=\frac{i}{4\pi^{2}}\Big(\frac{\lambda}{\sinh\lambda}\Big)^{1/2}\Big(\frac{\gamma}{\sin\gamma}\Big)^{1/2}\frac{1}{\lambda^{2}+\gamma^{2}},
\end{equation}
where we have suppressed the `$i\epsilon$' for compactness.

From Eq.(\ref{eq:Gdiv}) and Eq.(\ref{eq:ghartlehawking}), it is straightforward to calculate that
\begin{align}
W_{\textrm{H}}&=-8\pi^2 i(G_{\textrm{H}}-G_{\textrm{div}})\\
&= -\frac{1}{240}(\lambda^2+\gamma^2) + \mathcal{O}\bigl(\sigma^{3/2}\bigr)\\
&=-\frac{1}{240}\sigma^{;a}\sigma_{;a}+ \mathcal{O}\bigl(\sigma^{3/2}\bigr).
\end{align}
Comparing with the Taylor expansion (\ref{eq:wTaylor}), we can read off
\begin{align}
w_{H}(x)=0,\quad \varpi_{H}(x)^{a}{}_{b}=-\frac{1}{120}\delta^{a}{}_{b},
\end{align}
and hence the renormalized stress-energy tensor for the field in the Hartle-Hawking state is
\begin{equation}
\langle H|\hat{T}^{a}{}_{b}|H\rangle_{\textrm{ren}}=\frac{1}{2880\pi^2} \delta^{a}{}_{b}.
\end{equation}

\begin{figure*}
\centering
\includegraphics[width=16cm]{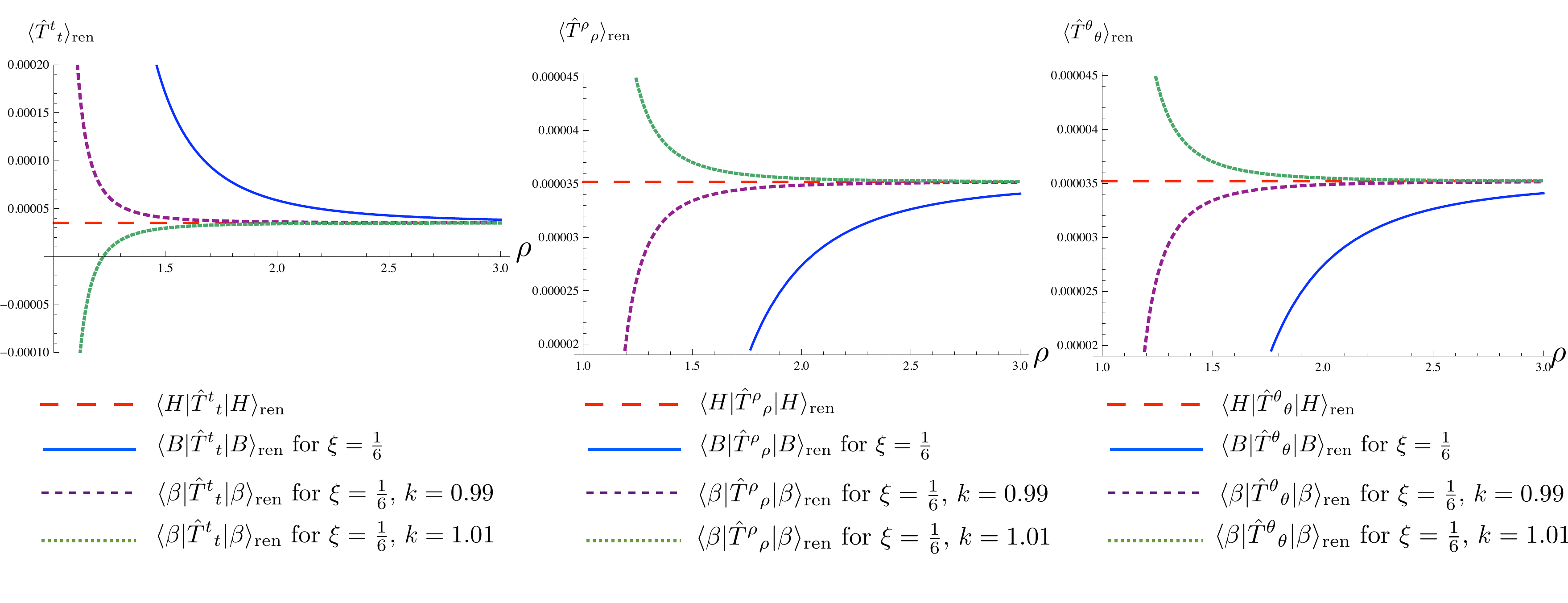}
\caption{{Plots of the renormalized stress-energy tensor components for the conformally coupled field for the various quantum states.} }
\label{fig:plotTrenConformal}
\end{figure*} 
\begin{figure*}
\centering
\includegraphics[width=16cm]{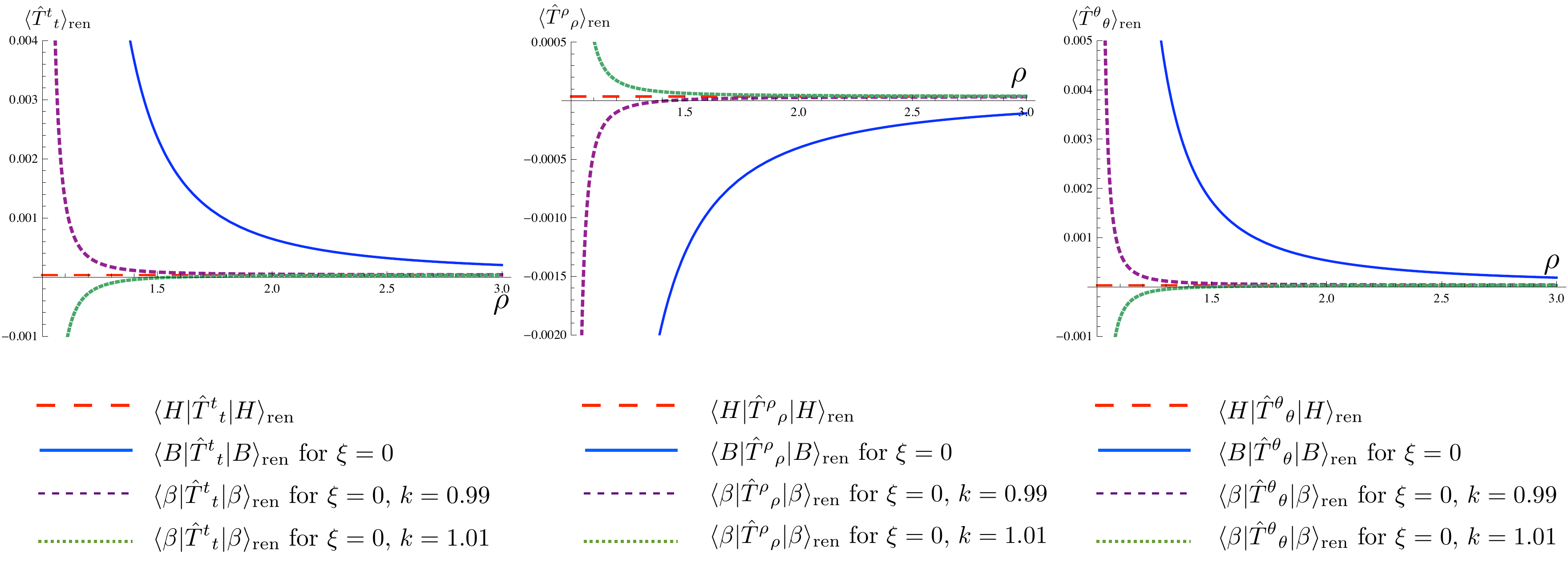}
\caption{{Plots of the renormalized stress-energy tensor components for the minimally coupled field for the various quantum states. } }
\label{fig:plotTrenMinimal}
\end{figure*} 

To compute $W$ for the Boulware vacuum, we note that
\begin{align}
W_{\textrm{H}}-W_{\textrm{B}}&=-8\pi^2 i(G_{\textrm{H}}-G_{\textrm{B}})\\
&=\frac{1}{6(\rho^2-1)^{1/2}({\rho'}^2-1)^{1/2}}\nonumber\\
&\quad\times\left(1-\frac{1}{60}(3\Delta t^2+11\eta^2) + \mathcal{O}\bigl(\sigma^{3/2}\bigr)\right),
\end{align}
which implies
\begin{align}
W_{B}=-\frac{1}{6(\rho^{2}-1)}-\frac{\rho\,\Delta\rho}{6(\rho^{2}-1)^{2}}-\frac{60\rho^{2}+19}{360(\rho^{2}-1)^{3}}\Delta\rho^{2}\nonumber\\
+\frac{1}{120(\rho^{2}-1)}\Delta t^{2}-\frac{3\rho^{4}-6\rho^{2}-19}{720(\rho^{2}-1)^{2}}\gamma^{2}-\frac{1}{240}\lambda^{2}\nonumber\\
+\,\mathcal{O}(\sigma^{3/2}).
\end{align}
Now noting that
\begin{align}
\sigma^{;t}&=  \Delta t - \frac{\rho}{\rho^2-1} \Delta t \Delta \rho + \mathcal{O}\bigl(\Delta x^{3}\bigr)\\
\sigma^{;\rho}&=  \Delta \rho -\frac{1}{2}\rho(\rho^2-1) \Delta t^2+ \frac{\rho}{2(\rho^2-1)} \Delta \rho^2 +\,\mathcal{O}\bigl(\Delta x^3\bigr)
\end{align}
which may be inverted to give
\begin{align}
\Delta \rho &=  \sigma^{;\rho} +\frac{1}{2}\rho(\rho^2-1) (\sigma^{;t})^2- \frac{\rho}{2(\rho^2-1)} (\sigma^{;\rho})^2 +\mathcal{O}\bigl(\sigma^{3/2}\bigr).
\end{align}
we find
\begin{align}
W_{B}&=-\frac{1}{6(\rho^{2}-1)}-\frac{\rho}{6(\rho^{2}-1)^{2}}\sigma^{;\rho}+\frac{\rho^{4}-22\rho^{2}+3}{240(\rho^{2}-1)}(\sigma^{;t})^{2}\nonumber\\
&-\frac{3\rho^{4}+54\rho^{2}+41}{720(\rho^{2}-1)^{3}}(\sigma^{;\rho})^{2}-\frac{3\rho^{4}-6\rho^{2}-19}{720(\rho^{2}-1)^{2}}g_{\alpha\beta}\sigma^{;\alpha}\sigma^{;\beta}\nonumber\\
&\qquad\qquad\qquad\qquad\qquad+\mathcal{O}(\sigma^{3/2}),
\end{align}
where $\alpha$, $\beta$ range over the angular coordinates. We can now read off
\begin{align}
\label{eq:wboulware}
w_B &= - \frac{1}{6(\rho^2-1)}\\
\varpi_B{}^{t}{}_{t}&=-\frac{\rho^4-22\rho^2+3}{120(\rho^2-1)^2}\\
\varpi_B{}^{\rho}{}_{\rho}&=-\frac{3\rho^4+54\rho^2+41}{360(\rho^2-1)^2}\\
\varpi_B{}^{\theta}{}_{\theta}&=\varpi_B{}^{\phi}{}_{\phi}=-\frac{3\rho^4-6\rho^2-19}{360(\rho^2-1)^2} .
\end{align}
which can be seen to satisfy the constraint Eq.~(\ref{eq:wconstraint}).

Applying the above results in Eq.~(\ref{eq:Trenw}), we find that the renormalized stress-energy tensor for the field in the Boulware vacuum is
\begin{align}
\langle B|\hat{T}^t{}_{t}|B\rangle_{\textrm{ren}}&=\frac{1}{2880\pi^2} -\frac{(\xi-\tfrac{1}{6})}{16\pi^{2}(\rho^{2}-1)}-\frac{(\xi-\tfrac{11}{60})}{8\pi^{2}(\rho^{2}-1)^{2}}
,\\
\langle B|\hat{T}^\rho{}_{\rho}|B\rangle_{\textrm{ren}}&=\frac{1}{2880\pi^2} +\frac{(\xi-\tfrac{1}{6})}{16\pi^{2}(\rho^{2}-1)}+\frac{(\xi-\tfrac{11}{60})}{24\pi^{2}(\rho^{2}-1)^{2}},\\
\langle B|\hat{T}^\theta{}_{\theta}|B\rangle_{\textrm{ren}}&=\langle B|\hat{T}^\phi{}_{\phi}|B\rangle_{\textrm{ren}}\nonumber\\
&=\frac{1}{2880\pi^2} -\frac{(\xi-\tfrac{1}{6})}{16\pi^{2}(\rho^{2}-1)}-\frac{(\xi-\tfrac{19}{120})}{12\pi^{2}(\rho^{2}-1)^{2}}.
\end{align}

A similar analysis for the Green's function for the field at an arbitrary temperature $T=1/\beta=k/2\pi$ reveals
\begin{align}
w_{\beta}&=w_{B}+\frac{k^{2}}{6(\rho^{2}-1)}\\
\varpi_{\beta}{}^{t}{}_{t}&=\varpi_{B}{}^{t}{}_{t}-\frac{k^{2}(10\rho^{2}-k^{2})}{60(\rho^{2}-1)^{2}}\\
\varpi_{\beta}{}^{\rho}{}_{\rho}&=\varpi_{B}{}^{\rho}{}_{\rho}+\frac{k^{2}(30\rho^{2}+20-k^{2})}{180(\rho^{2}-1)^{2}}\\
\varpi_{\beta}{}^{\theta}{}_{\theta}&=\varpi_{\beta}{}^{\phi}{}_{\phi}=\varpi_{B}{}^{\theta}{}_{\theta}-\frac{k^{2}(10+k^{2})}{180(\rho^{2}-1)^{2}},
\end{align}
so that 
\begin{equation}
\lim_{T\rightarrow 0}W_{\beta}(x,x')=W_{B}(x,x'),
\end{equation}
 as expected. Finally, from Eq.(\ref{eq:Trenw}), we obtain the following expressions for the renormalized stress-energy tensor for the field at an arbitrary temperature $T$,
\begin{align}
\label{eq:Trenbeta}
\langle \beta|\hat{T}^t{}_{t}|\beta\rangle_{\textrm{ren}}&=\langle B|\hat{T}^t{}_{t}|B\rangle_{\textrm{ren}}+\frac{k^{2}(\xi-\tfrac{1}{6})(\rho^{2}+1)}{16\pi^{2}(\rho^{2}-1)^{2}}\nonumber\\
&\qquad\qquad\qquad-\frac{k^{4}}{480\pi^{2}(\rho^{2}-1)^{2}}
,\\
\langle \beta|\hat{T}^\rho{}_{\rho}|\beta\rangle_{\textrm{ren}}&=\langle B|\hat{T}^\rho{}_{\rho}|B\rangle_{\textrm{ren}}-\frac{k^{2}(\xi-\tfrac{1}{6})(3\rho^{2}-1)}{48\pi^{2}(\rho^{2}-1)^{2}}\nonumber\\
&\qquad\qquad\qquad+\frac{k^{4}}{1440\pi^{2}(\rho^{2}-1)^{2}},\\
\langle \beta|\hat{T}^\theta{}_{\theta}|\beta\rangle_{\textrm{ren}}&=\langle B|\hat{T}^\theta{}_{\theta}|B\rangle_{\textrm{ren}}+\frac{k^{2}(\xi-\tfrac{1}{6})(3\rho^{2}+1)}{48\pi^{2}(\rho^{2}-1)^{2}}\nonumber\\
&\qquad\qquad\qquad+\frac{k^{4}}{1440\pi^{2}(\rho^{2}-1)^{2}}\nonumber\\
&=\langle \beta|\hat{T}^\phi{}_{\phi}|\beta\rangle_{\textrm{ren}}.
\end{align}

In Fig.~\ref{fig:plotTrenConformal}, we have plotted the renormalized stress-energy tensor for the conformally coupled field for the three distinct quantum states, where for the thermal state we include a graph for the field with a temperature greater than the Hawking temperature and a graph where the field is less than the Hawking temperature. The temperature $T=k/2\pi$ can be increased continuously from zero, which corresponds to the Boulware state. We see that for $T<T_{H}$ ($k<1$), we have $\langle \beta|\hat{T}^{t}{}_{t}|\beta\rangle-\langle H|\hat{T}^{t}{}_{t}|H\rangle>0$, while for $T>T_{H}$, this difference becomes positive, reflecting the fact that the black hole is a more efficient attractor of a particle with a temperature greater than the Hawking temperature. From the expressions (\ref{eq:Trenbeta}), we see that the contribution from the temperature is $\mathcal{O}(T^{4})$, which dominates for $k>1$, but which is suppressed for $k<<1$. The components for the Boulware and thermal states are asymptotically equal to the Hartle-Hawking state so that all observers at fixed large radius will measure a thermal spectrum at the Hawking temperature, regardless of the quantum state. As expected, with the exception of the Hartle-Hawking state, the stress-energy tensor is singular at the horizon.

In Fig.~\ref{fig:plotTrenMinimal}, we plot the renormalized stress-energy tensor for the minimally coupled field. We have most of the same qualitative features as in the conformally coupled case, with a few minor exceptions. The temperature dependance of the stress-energy tensor components is now stronger for the field at a temperature less than the Hartle-Hawking temperature, compared with the conformally coupled case.  This arises as a result of the $\mathcal{O}(T^{2})$ terms in (\ref{eq:Trenbeta}). For $T>T_{H}$ ($k>1$), the $\mathcal{O}(T^{4})$ will still dominate yielding a behavior qualitatively similar to the conformally coupled case. The angular components of the stress-energy tensor exhibit a very different behavior for the conformally and minimally coupled cases, with the difference between the components in the Boulware or thermal states and the Hartle-Hawking state changes sign for minimally coupled field relative to the conformal field.

\acknowledgements
We are most grateful to Marc Casals for numerous enlightening conversations.

\appendix*
\section{Green's Function on the Dimensionally Reduced Bertotti Robinson Space-Time}
In this Appendix, we derive a closed-form expression for the three-dimensional Green's function on the dimensionally reduced Bertotti-Robinson space-time, obtained by factoring out the temporal dependence by a Fourier transform. The method employed is a novel expansion about the zero frequency solution, similar in spirit to the method adopted by Copson \cite{Copson} to derive the electrostatic potential in the Schwarzschild space-time. It may seem surprising that such an expansion truly captures the global properties of the solution since it is well known that a Hadamard expansion in the geodesic distance cannot capture the global properties but only captures the singular behaviour of the Green's function \cite{Hadamard}. However, the crucial point is that our method is not an expansion about a purely geometrical quantity such as the geodesic distance but rather is an expansion in the zero-frequency solution, which already encodes the boundary condition of the problem. Hence, the expansion need only capture the``potentialness'' of the problem.

We work with the metric of Eq.(\ref{eq:brmetric}) with $L_{p}^{2}Q^{2}=1$. The Green's function for a massless scalar field on this space-time satisfies
\begin{align}
&\Big[-\frac{1}{(\rho^{2}-1)}\frac{\partial^{2}}{\partial t^{2}}+\frac{\partial}{\partial\rho}\Big((\rho^{2}-1)\frac{\partial}{\partial \rho}\Big)+\frac{1}{\sin\theta}\frac{\partial}{\partial\theta}\Big(\!\sin\theta\frac{\partial}{\partial\theta}\Big)\nonumber\\
&\qquad
+\frac{1}{\sin^{2}\theta}\frac{\partial^{2}}{\partial\phi^{2}}\Big]G(x,x')
=\frac{1}{ \sin\theta}\delta(x-x').
\end{align}
To obtain a unique solution to the wave equation for a specific set of boundary conditions, we euclideanize the metric by analytically continuing the temporal coordinate $t=-i \tau$. Then the relationship between the Euclidean and Feynman Green's functions is given by \cite{BD}
\begin{equation}
\label{eq:euclideantofeynman}
G_{F}(t,\textbf{x}; t', \textbf{x}')=-i G_{E}(i\tau, \textbf{x}; i\tau, \textbf{x}').
\end{equation}
It is straight-forward to show that by a separation of variables, the Euclidean Green's function may be written in the following form:
\begin{align}
&G_{E}(x,x')=\nonumber\\
&\frac{1}{4\pi^{2}}\int_{0}^{\infty} d\omega\> \cos\omega(\tau-\tau')  \sum_{l=0}^{\infty}(2 l+1)P_{l}(\cos\gamma)\chi_{\omega l}(\rho,\rho')
\end{align}
where $\cos\gamma=\cos\theta\cos\theta'+\sin\theta\sin\theta'\cos(\phi-\phi')$ and $\chi_{\omega l}$ satisfies the inhomogeneous equation
\begin{align}
\Big[\frac{d}{d\rho}\Big((\rho^{2}-1)\frac{d}{d\rho}\Big)-\frac{\omega^{2}}{(\rho^{2}-1)}-l(l+1)\Big]\chi_{\omega l}(\rho,\rho')\nonumber\\
=-\delta(\rho-\rho').
\end{align}
The solutions of the corresponding homogeneous equation are the Associated Legendre functions \cite{gradriz} of non-integer order where the particular choice depends on the boundary conditions or the quantum state for the problem under consideration. Requiring regularity at the horizon $\rho=1$ and vanishing at infinity corresponds to taking the Associated Legendre function of the first kind, $P_{l}^{-|\omega|}(\rho)$, to be the inner solution and the Associated Legendre function of the second kind, $Q^{\pm |\omega|}_{l}(\rho)$ to be the outer solution, where the product of homogeneous solutions is normalized by the Wronskian. The Wronskian takes a particularly simple form for the set of solutions $\big\{P^{-|\omega|}_{l}, Q^{|\omega|}_{l}\big\}$,
\begin{align}
W\big[P^{-|\omega|}_{l}, Q^{|\omega|}_{l}\big]&=P^{-|\omega|}_{l}(\rho)\frac{d Q^{|\omega|}_{l}(\rho)}{d \rho}-\frac{d P^{-|\omega|}_{l}(\rho)}{d \rho} Q^{|\omega|}_{l}(\rho)\nonumber\\
&=-\frac{e^{i |\omega| \pi}}{(\rho^{2}-1)}.
\end{align}
The Green's function then takes the form
\begin{align}
\label{eq:gmodesum}
G_{E}(x,x')=\frac{1}{4\pi^{2}}\int_{0}^{\infty} d\omega \>\cos\omega(\tau-\tau')\sum_{l=0}^{\infty}(2l+1)\nonumber\\
P_{l}(\cos\gamma)e^{-i\omega \pi}P_{l}^{-\omega}(\rho_{<})Q_{l}^{\omega}(\rho_{>})
\end{align}
where we can drop the absolute value since $\omega$ only runs over non-negative values. 

We will now show that the $l$-sum in this mode sum is itself a three-dimensional Green's function on a dimensionally reduced Bertotti-Robinson space-time. Writing the four-dimensional Green's function as
\begin{equation}
\label{eq:greensreduced}
G_{E}(x,x')=\frac{1}{4\pi^{2}}\int_{0}^{\infty} d\omega\>\cos\omega(\tau-\tau') G_{\omega}(\textbf{x},\textbf{x}')
\end{equation}
and substituting into our wave-equation, it is easy to see that $G_{\omega}(\textbf{x},\textbf{x}')$ satisfies
\begin{align}
\Big[\frac{\partial}{\partial\rho}\Big((\rho^{2}-1)\frac{\partial}{\partial\rho}\Big)+\frac{1}{\sin\theta}\frac{\partial}{\partial\theta}\Big(\sin\theta \frac{\partial}{\partial\theta}\Big)+\frac{1}{\sin^{2}\theta}\frac{\partial^{2}}{\partial\phi^{2}}\nonumber\\-\frac{\omega^{2}}{(\rho^{2}-1)}\Big]G_{\omega}(\textbf{x},\textbf{x}')
=-\frac{1}{\sin\theta}\delta(\textbf{x}-\textbf{x}').
\end{align}
Dividing across by $(\rho^{2}-1)$ gives the Helmholtz equation
\begin{equation}
\label{eq:waveeqnreducedbr}
\Big[\nabla^{2}-\frac{\omega^{2}}{(\rho^{2}-1)^{2}}\Big]G_{\omega}(\textbf{x},\textbf{x}')=-\tilde{g}^{-1/2}\delta(\textbf{x}-\textbf{x}'),
\end{equation}
where $\nabla^{2}$ and $\tilde{g}$ are the Laplacian and metric determinant, respectively, on the three-metric
\begin{equation}
\label{eq:metricreducedbr}
ds_{3}^{2}=\tilde{g}_{\alpha\beta}dx^{\alpha}dx^{\beta}=d\rho^{2}+(\rho^{2}-1)d\theta^{2}+(\rho^{2}-1)\sin^{2}\theta d\phi^{2}.
\end{equation}
Comparing the mode-sum expression Eq.(\ref{eq:gmodesum}) with Eq.(\ref{eq:greensreduced}) yields
\begin{equation}
\label{eq:greensfnmodesumreducedbr}
G_{\omega}(\textbf{x},\textbf{x}')=\sum_{l=0}^{\infty}(2l+1)P_{l}(\cos\gamma)e^{-i\omega \pi}P_{l}^{-\omega}(\rho_{<})Q_{l}^{\omega}(\rho_{>}).
\end{equation}

We wish to obtain a closed-form solution for $G_{\omega}(\textbf{x}, \textbf{x}')$. Since the sum on the right-hand side of Eq.(\ref{eq:greensfnmodesumreducedbr}) is not known, except for integer values of $\omega$ \cite{Candelas:1984pg}, we must employ an alternative method for obtaining the closed-form solution. Moreover, if such a closed-form representation can be obtained, we have derived a new summation formula for the product of associated Legendre functions of arbitrary order appearing in Eq.(\ref{eq:greensfnmodesumreducedbr}).

Another way of writing the Green's function of a wave equation in arbitrary dimensions is in Hadamard form \cite{Decanini:2008}, which is an expansion in Synge's world function $\sigma(x,x')$ which is half the square of the geodesic distance between $x$ and $x'$. This expansion is purely geometrical and therefore fails to capture the boundary conditions. Rather than expanding about $\sigma$, we could try an expansion about the zero-potential solution in Eq.(\ref{eq:greensfnmodesumreducedbr}) since, for $\omega=0$, a closed-form solution is known \cite{Candelas:1984pg},
\begin{equation}
G_{0}(\textbf{x},\textbf{x}')=\frac{1}{R^{1/2}}
\end{equation}
where
\begin{equation}
R=\rho^{2}+\rho'^{2}-2\rho\rho'\cos\gamma-\sin^{2}\gamma.
\end{equation}
We shall show that this approach does give the globally valid closed-form solution.

The problem may be simplified by assuming the \textit{ansatz}
\begin{equation}
G_{\omega}(\textbf{x},\textbf{x}')=\frac{e^{-\omega \,\,U(\textbf{x},\textbf{x}')}}{R^{1/2}}
\end{equation}
and substituting into the Helmholtz equation to get
\begin{align}
-\omega\nabla^{2}U+\omega^{2}\tilde{g}^{\alpha \beta}\frac{\partial U}{\partial x^{\alpha}}\frac{\partial U}{\partial x^{\beta}}+\omega\frac{\tilde{g}^{\alpha \beta}}{R}\frac{\partial U}{\partial x^{\alpha}}\frac{\partial R}{\partial x^{\beta}}-\frac{\omega^{2}}{(\rho^{2}-1)^{2}}\nonumber\\
=0.
\end{align}
Equating powers of $\omega$, we arrive at the following equations which must be simultaneously satisfied
\begin{align}
\label{eq:U2equation}
\tilde{g}^{\alpha \beta}\frac{\partial U}{\partial x^{\alpha}}\frac{\partial U}{\partial x^{\beta}}-\frac{1}{(\rho^{2}-1)^{2}}&=0 \\
\label{eq:gradUequation}
\nabla^{2}U-\frac{\tilde{g}^{\alpha \beta}}{R}\frac{\partial U}{\partial x^{\alpha}}\frac{\partial R}{\partial x^{\beta}}&=0.
\end{align}
The first of these equations, though non-linear, turns out to be easier to solve. We look for a series solution of the form
\begin{equation}
U(\textbf{x},\textbf{x}')=\sum_{k=0}^{\infty} a_{k}(\rho, R) R^{k+1/2},
\end{equation}
where the dependence of the $a_{k}$ coefficients as functions of $\rho$ and $R$ only is a consequence of the spherical symmetry. Substituting this series solution into Eq.(\ref{eq:U2equation}) and eliminating any $\cos\gamma$ dependence in favour of $\rho$ and $R$ using
\begin{align}
\label{eq:cosrelations1}
\cos\gamma&=\rho\rho'-(R+(\rho^{2}-1)(\rho'^{2}-1))^{1/2} \nonumber\\
&=\rho\rho'-\sum_{m=0}^{\infty} {\tfrac{1}{2}\choose m} \frac{R^{m}}{(\rho^{2}-1)^{m-1/2}(\rho'^{2}-1)^{m-1/2}},
\end{align}
we obtain
\begin{widetext}
\begin{align}
\label{eq:seriesU2etaR}
\tilde{g}^{\alpha \beta} \frac{\partial U}{\partial x^{\alpha}}\frac{\partial U}{\partial x^{\beta}}-\frac{1}{(\rho^{2}-1)^{2}}&=\sum_{k=0}^{\infty}\sum_{s=0}^{\infty}\Big(\frac{\partial a_{k}}{\partial \rho}\frac{\partial a_{s}}{\partial \rho} R^{k+s+1}-\rho(\rho'^{2}-1)(2 s+1)a_{s} \frac{\partial a_{k}}{\partial \rho} R^{k+s}-\rho(\rho'^{2}-1)(2 k+1)a_{k} \frac{\partial a_{s}}{\partial \rho} R^{k+s}\nonumber\\
&-(2k+1)(2s+1)a_{s}a_{k}\frac{R^{k+s+1}}{(\rho^{2}-1)}+(2k+1)(2s+1)a_{s}a_{k}(1-2\rho^{2})\frac{(\rho'^{2}-1)}{(\rho^{2}-1)}R^{k+s} \nonumber\\
&+(2 s+1)a_{s}\frac{\partial a_{k}}{\partial\rho}\rho'\sum_{m=0}^{\infty}{\tfrac{1}{2}\choose m}\frac{R^{k+s+m}}{(\rho^{2}-1)^{m-1/2}(\rho'^{2}-1)^{m-1/2}} \nonumber\\
&+(2 k+1)a_{k}\frac{\partial a_{s}}{\partial\rho}\rho'\sum_{m=0}^{\infty}{\tfrac{1}{2}\choose m}\frac{R^{k+s+m}}{(\rho^{2}-1)^{m-1/2}(\rho'^{2}-1)^{m+1/2}} \nonumber\\
&+(2k+1)(2s+1)a_{s}a_{k} 2\rho\rho'\sum_{m=0}^{\infty}{\tfrac{1}{2}\choose m}\frac{R^{k+s+m}}{(\rho^{2}-1)^{m+1/2}(\rho'^{2}-1)^{m-1/2}} \Big)-\frac{1}{(\rho^{2}-1)^{2}}=0.
\end{align}
\end{widetext}
Equating terms proportional to $R^{0}$, we get
\begin{align}
\frac{d a_{0}^{2}}{d\rho} (\rho'^{2}-1)^{1/2}(\rho'(\rho^{2}-1)^{1/2}-\rho(\rho'^{2}-1)^{1/2})\nonumber\\
+a_{0}^{2}\frac{(\rho'^{2}-1)}{(\rho^{2}-1)}\Big((1-2\rho^{2})+2\rho\rho'\frac{(\rho^{2}-1)^{1/2}}{(\rho'^{2}-1)^{1/2}}\Big)\nonumber\\
-\frac{1}{(\rho^{2}-1)^{2}}=0
\end{align}
This is a simple first order differential equation which may be solved using an integrating factor. Dividing across by the coefficient of $a_{0}^{2}$, then the integrating factor is given by
\begin{align}
I(\rho)=\exp\Bigl(\int P(\rho) \,d\rho \Bigr)
\end{align}
where
\begin{align}
P(\rho)=\frac{2\rho\rho'(\rho^{2}-1)^{1/2}+(1-2\rho^{2})(\rho'^{2}-1)^{1/2}}{(\rho^{2}-1)^{1/2}(\rho'(\rho^{2}-1)^{1/2}-\rho(\rho'^{2}-1)^{1/2})}.
\end{align}
The integral here may be performed to give
\begin{equation}
I(\rho)=(\rho^{2}-1)^{1/2}(\rho'(\rho^{2}-1)^{1/2}-\rho(\rho'^{2}-1)^{1/2}).
\end{equation}
Multiplying across by the integration factor we may re-write the equation in the succinct form,
\begin{equation}
\frac{d}{d\rho}\big(a_{0}^{2}I(\rho)\big)=\frac{1}{(\rho^{2}-1)^{3/2}(\rho'^{2}-1)^{1/2}}.
\end{equation}
Integrating both sides and dividing across by $I(\rho)$, we get
\begin{equation}
a_{0}^{2}=\frac{-\rho(\rho'^{2}-1)^{1/2}+f(\rho')(\rho^{2}-1)^{1/2}(\rho'^{2}-1)}{(\rho^{2}-1)(\rho'^{2}-1)[\rho'(\rho^{2}-1)^{1/2}-\rho(\rho'^{2}-1)^{1/2}]}
\end{equation}
where $f(\rho')$ is a function of $\rho'$ only (arising as the integration ``constant'' in the integral with respect to $\rho$) and is chosen in such a way that $a_{0}$ be symmetric in $\rho$ and $\rho'$. This condition fixes $f(\rho')$ to be
\begin{equation}
f(\rho')=\frac{\rho'}{(\rho'^{2}-1)}
\end{equation}
so that
\begin{equation}
a_{0}=\frac{1}{(\rho^{2}-1)^{1/2}(\rho'^{2}-1)^{1/2}}.
\end{equation}

Similarly, equating terms that are linear in $R$, we obtain
\begin{align}
\{2(\rho'^{2}-1)^{1/2}(\rho'(\rho^{2}-1)^{1/2}-\rho(\rho'^{2}-1)^{1/2})\}\frac{d a_{1}}{d\rho}\nonumber\\
+\frac{6(\rho'^{2}-1)^{1/2}}{(\rho^{2}-1)}\big[\rho\rho'(\rho^{2}-1)^{1/2}+(1-\rho^{2})(\rho'^{2}-1)^{1/2}\big]a_{1}\nonumber\\
+\frac{a_{0}}{(\rho^{2}-1)^{2}}=0
\end{align}
where we have used the fact that
\begin{equation}
\frac{d a_{0}}{d\rho}=-\frac{\rho}{(\rho^{2}-1)}a_{0}.
\end{equation}
Dividing across by the coefficient of the first term, we may re-write as
\begin{align}
0=\frac{da_{1}}{d\rho}+\frac{3a_{1}[\rho\,\rho'-(\rho^{2}-1)^{1/2}(\rho'^{2}-1)^{1/2}]}{(\rho^{2}-1)^{1/2}[\rho'(\rho^{2}-1)^{1/2}-\rho(\rho'^{2}-1)^{1/2}]} \nonumber\\
+\frac{a_{0}}{2(\rho'^{2}-1)^{1/2}(\rho^{2}-1)^{2}[\rho'(\rho^{2}-1)^{1/2}-\rho(\rho'^{2}-1)^{1/2}]}
\end{align}
With an integrating factor
\begin{equation}
I(\rho)=\frac{1}{\rho'^{3}}\big[\rho'(\rho^{2}-1)^{1/2}-\rho(\rho'^{2}-1)^{1/2}\big]^{3}
\end{equation}
the equation for $a_{1}$ may be solved in an analogous way to give
\begin{equation}
a_{1}=-\frac{1}{6}\frac{1}{(\rho^{2}-1)^{3/2}(\rho'^{2}-1)^{3/2}}.
\end{equation}

In fact all of the $a_{k}$'s may be obtained in this way (though the algebra becomes increasingly cumbersome). The first few terms are
\begin{align}
a_{0}&=\frac{1}{(\rho^{2}-1)^{1/2}(\rho'^{2}-1)^{1/2}} \nonumber\\
a_{1}&=-\frac{1}{6}\frac{1}{(\rho^{2}-1)^{3/2}(\rho'^{2}-1)^{3/2}} \nonumber\\
a_{2}&= \frac{3}{40}\frac{1}{(\rho^{2}-1)^{5/2}(\rho'^{2}-1)^{5/2}} \nonumber\\
a_{3}&=- \frac{5}{112}\frac{1}{(\rho^{2}-1)^{7/2}(\rho'^{2}-1)^{7/2}} \nonumber\\
a_{4}&= \frac{35}{1152}\frac{1}{(\rho^{2}-1)^{9/2}(\rho'^{2}-1)^{9/2}}\qquad \textrm{etc.}
\end{align}
The coefficients here may be shown by induction to correspond to the series
\begin{align}
\cosh^{-1}\big((x+1)^{1/2}\big)&=x^{1/2}-\tfrac{1}{6}x^{3/2}+\tfrac{3}{40}x^{5/2}-\tfrac{5}{112}x^{7/2}\nonumber\\
&\qquad+\dots
\end{align}
Therefore, we have that $U(\textbf{x},\textbf{x}')$ is given by
\begin{align}
U(\textbf{x},\textbf{x}')&=\cosh^{-1}\Big[\Big(\frac{R}{(\rho^{2}-1)(\rho'^{2}-1)}+1\Big)^{1/2}\Big]\nonumber\\
&=\cosh^{-1}\Big[\frac{\rho\rho'-\cos\gamma}{(\rho^{2}-1)^{1/2}(\rho'^{2}-1)^{1/2}}\Big].
\end{align}
The closed-form expression for the Green's function on the dimensionally reduced Bertotti-Robinson space-time is therefore given by
\begin{align}
\label{eq:gomega}
G_{\omega}(\textbf{x},\textbf{x}')=\frac{e^{-\omega \eta}}{R^{1/2}}
\end{align}
where
\begin{align}
\label{eq:chidefn}
\cosh\eta=\frac{\rho\rho'-\cos\gamma}{(\rho^{2}-1)^{1/2}(\rho'^{2}-1)^{1/2}}.
\end{align}

Equating Eq.(\ref{eq:gomega}) with its equivalent mode-sum expression yields the following new summation formula for Associated Legendre functions of arbitrary order:
\begin{align}
\label{eq:legendresumomega}
\sum_{l=0}^{\infty}(2 l+1) P_{l}(\cos\gamma) e^{-i \omega \pi} P_{l}^{-\omega}(\rho_{<})Q_{l}^{\omega}(\rho_{>})\nonumber\\
=\frac{e^{-\omega \eta}}{(\rho^{2}+\rho'^{2}-2\rho\rho'\cos\gamma-\sin^{2}\gamma)^{1/2}}.
\end{align}

It is worth noting that, since the derivation of the closed-form expression did not impose any restrictions on $\omega$, this result is valid for arbitrary complex $\omega$. In the case where $\omega=n$, an integer, then the left-hand side is invariant under $n\rightarrow -n$ and so we must choose $\omega=|n|$ on the right-hand side to reflect this. This summation formula was crucial in obtaining the closed-form representation of the Green's function in the Boulware vacuum and for a field at an arbitrary temperature in Sec.~\ref{sec:propagators}.

There are also contexts in which it is useful to be able to write a product of associated Legendre functions as an integral whereby all of the dependence on the order of the Legendre functions is contained in an exponential function. For example, multiplying on both sides by $P_{l'}(\cos\gamma)$ and integrating with respect to $\cos\gamma$, we arrive at
\begin{align}
\label{eq:brproduct}
&e^{-i\omega\pi} P_{l}^{-\omega}(\rho_{<}) Q_{l}^{\omega}(\rho_{>})\nonumber\\
&=\frac{1}{2}\int_{-1}^{1}d(\cos\gamma)\, \frac{e^{-\omega \eta} P_{l}(\cos\gamma)}{(\rho^{2}+\rho'^{2}-2\rho\rho'\cos\gamma-\sin^{2}\gamma)^{1/2}}
\end{align}
where we used the standard normalization for the Legendre functions \cite{gradriz}
\begin{equation}
\int_{-1}^{1} P_{l}(x) P_{l'}(x) dx = \frac{2}{(2l+1)}\delta_{l l'}.
\end{equation}
Such integral representations are useful since the only dependence on the non-integer order is in the argument of the exponential, which is typically easier to sum or integrate in a mode-sum expression. In Ref.~\cite{OttewillTaylor3}, we have used result (\ref{eq:brproduct}) to obtain a closed-form solution for the retarded Green's function for a static scalar particle in the Kerr space-time and hence obtain the self-force for such a particle.

A similar but more succinct form may be obtained by changing the integration variable to $\eta$,
\begin{align}
&e^{-i\omega\pi} P_{l}^{-\omega}(\rho_{<}) Q_{l}^{\omega}(\rho_{>})\nonumber\\
&=\frac{1}{2}\int_{\eta_{-}}^{\eta_{+}} d\eta\> e^{-\omega\eta} P_{l}(\rho\rho'-(\rho^{2}-1)^{1/2}(\rho'^{2}-1)^{1/2}\cosh\eta)
\end{align}
where
\begin{equation}
\eta_{\pm}=\cosh^{-1}\left[\frac{\rho \rho'\pm1}{(\rho^{2}-1)^{1/2}(\rho'^{2}-1)^{1/2}}\right].
\end{equation}

Yet another form may be obtained by further changing the radial variable
$\rho=\cosh\xi$
which decouples the ``primed'' coordinates from the ``unprimed'' coordinates in the integration limits, yielding
\begin{align}
\label{eq:legendreproduct}
&e^{-i\omega\pi} P_{l}^{-\omega}(\rho_{<}) Q_{l}^{\omega}(\rho_{>})\nonumber\\
&=\frac{1}{2}\int_{\eta_{-}}^{\eta_{+}} d\eta\> e^{-\omega\eta} P_{l}(\cosh\xi\cosh\xi'-\sinh\xi\sinh\xi'\cosh\eta)
\end{align}
where now the limits of integration are given by
\begin{equation}
\eta_{\pm}=-\log\left[\tanh\left(\frac{\xi_{<}}{2}\right)\right]\pm\log\left[\tanh\left(\frac{\xi_{>}}{2}\right)\right].
\end{equation}

\bibliographystyle{apsrev}
\bibliography{database}
\end{document}